\documentclass[preprint]{aastex}
\usepackage{natbib}

\begin{document}

\title{The [Fe/H] Dependence on the Ca~{\sc ii}-$M_V$ Relationship}
\author{Thomas Gomez$^1$, George Wallerstein$^2$, Elena Pancino$^3$}
\affil{$^1$Department of Astronomy, University of Texas, Austin, TX 78712}
\affil{$^2$Department of Astronomy, University of Washington, Seattle, WA 98915}
\affil{$^3$INAF -Osservatorio Astronomico, Via Ranzani 1,40127 Bologna, Italy}

\affil{gomezt@astro.as.utexas.edu, wall@astro.washington.edu, elena.pancino@oabo.inaf.it}
\date{\today}

\begin{abstract}
We examined the Wilson-Bappu effect, a relationship between the absolute magnitude of the star, $M_V$, and the logarithm of the Ca~{\sc ii} emission width, $W_0$, over the largest $M_V$ range to date, +13 to -5, covering M-dwarfs to type Ia supergiants. We used an extensive literature, the latest Hipparcos reduction, data from two globular clusters, and new observations from Apache Point Observatory to compile a sample that allowed us to study the effect of [Fe/H] on the Wilson-Bappu relationship. Our results include reporting the deviations from linearity and demonstrating that the Wilson-Bappu relationship is insensitive to metallicity.
\end{abstract}

\section{Introduction}

O.C. \citet*{1957ApJ...125..661W} published a remarkable linear correlation between the logarithm of the Ca~II K emission full width at half maximum ($\log~W_0$) and the absolute visual magnitude of the star, $M_V$.  They presented that this holds for non-variable cool stars below type F, regardless of spectral type, [Fe/H], or intensity of the emission line. This correlation is useful to derive an $M_V$ for cool stars too distant for accurate parallaxes. This relationship can also be found in other spectral lines, \citet{1989A&A...208..198A} were the first to publish a similar relationship between $M_V$ and the with of the Mg~{\sc ii} k line.

The Wilson-Bappu (hereafter WB) effect has aided in studies of variable stars like T Tauri stars as well as main sequence stars (see \citet{1965PASP...77..253K} and  \citet{baliunas}). The WB effect has been invaluable for our ability to understand chromospheres and heating, see seminal work by \citet{1973A&A....24...79R}, \citet{1974A&A....35...99N} and \citet{1979ApJ...228..509A}.  \citeauthor{1973A&A....24...79R} and \citeauthor{1974A&A....35...99N} derived a way to relate $W_0$ to stellar parameters: $W \sim g^{-2.0} T^{1.1} [Fe/H]^{0.0}$. \citeauthor{1979ApJ...228..509A} was able to establish that the WB effect is largely a result of the surface gravity of the star. The Ca~{\sc ii} line is so optically thick that the increase in width corresponds to an increase in column density in the chromosphere. This relationship between column density and $W_0$ is predicted to be insensitive to [Fe/H]. 

Accurate Hipparcos parallaxes were used for the first time \citet{1999PASP..111..335W} to study the WB effect for a large number of stars with accurate $M_V$. Their $\log~W_0$ were taken from the large list published by \citet{1976ApJ...205..823W}. \citet{1999PASP..111..335W} were the first to use Hipparcos parallaxes and noticed that their small sample of supergiant stars showed unusually large $\log~W_0$ and deviated from a linear trend.

\citet{1957ApJ...125..661W} foudn that the WB effect has no [Fe/H] dependence. However, they had relatively few stars compared to recent studies of the WB effect and there are few metal-poor stars in the solar neighborhood. A chromospheric study of metal-poor giants, \citep{1995AJ....110..405D},hereafter DS95, demonstrated that extremely metal-poor stars ([Fe/H] $<$ -2.0) have a diminished $W_0$ and did not follow the WB relationship previously established by \citet{1976ApJ...205..823W}. The field giants they studied were too far away to have accurate parallaxes, therefore they estimated their luminosity by fitting each star's colors to isochrones of M92, an extremely metal-poor globular cluster. \citet{2004A&A...413..343C} compared stars they studied in the globular cluster NGC2808 to the study of field stars by \citet{2003yCat..34010997P} and came to a similar conclusion as \citet{1995AJ....110..405D}, where more metal poor stars lie above the linear fit derived from more metal rich stars. However, their result was much more subtle then \citeauthor{1995AJ....110..405D}

We expand the work by \citet{1999PASP..111..335W} and DS95 by using the latest Hipparcos reduction \citep{2007A&A...474..653V} and constraining our sample to stars with parallaxes greater than 3 times the probable error $(\Pi/\sigma_\Pi \geq 3)$\footnote{When studying individual stars, using parallaxes that are less than 5 times their uncertainty can introduce large errors and unreliable results. For a substantial set of data from which the general behavior of the sample is to be understood, a less strict limit provides sufficient accuracy for the purpose at hand.} instead of $(\Pi/\sigma_\Pi \geq 5)$. We have included all available measurements of $W_0$ from the literature, additional metal-poor stars observed at the Apache Point Observatory, and Ca~{\sc ii} widths of stars from the globular clusters NGC2808 and $\omega$ Cen.

\section{The Database}
Parallax is the most reliable and direct method of deriving $M_V$. $Hipparcos$ constitute the largest sample and most reliable parallaxes, so we used $Hipparcos$ parallaxes \citep{2007A&A...474..653V} for stars that meet our criteria of $\Pi/\sigma_{\Pi} \geq 3$.  The \citeauthor{2007A&A...474..653V} reduction of $Hipparcos$ data reduced the uncertainty in the parallaxes by a factor of 2 from \citet{1997ESASP1200.....P}, allowing us to study many more stars with acceptable parallaxes. 

To investigate the WB effect and its dependence on [Fe/H], we have assembled data from a list of previously published Ca~{\sc ii} emission widths, $W_0$. $W_0$ is defined as the width of the line, $W$, having the instrumental profile removed. See figure \ref{profile} for an example of how to measure $W$. $W_0$ values came from the catalogs by \citet{1980AcA....30..453G}, \citet{1994yCat..41080611L} as well as \citet{1983ApJS...53..815Z}, \citet{1999PASP..111..335W}, and \citet{2003yCat..34010997P}. For the M-dwarfs in our sample, we used Ca~{\sc ii} K full width at half maximum (FWHM) measurements from \citet{2006PASP..118..617R}. The sample of $W_0$ values still lacks a sufficient number of metal-poor stars, so we included the data by DS95 (shown in Table \ref{dupree}). In the case of a star with multiple observations, the most recent value of $W_0$ was used. For consistency, we used only values of $W_0$. Most of these publications used $W_0$, but some published values of $W$. The instrumental profiles of the spectrographs used were quoted in the paper and we subtracted the value from their published $W$ to obtain $W_0$. We compared the different references and methods to look for systematic differences in $W_0$ that would affect our study. There was no difference in the Wilson-Bappu effect between the different references.
We chose to use the catalog by \citet{2001yCat.3221....0C} for our values of [Fe/H] for consistency.

To study other main sequence stars, we analyzed the spectra from the McDonald Observatory archive \citep{2004A&A...420..183A} and measured $W_0$ ourselves. The spectra from McDonald were already reduced and flux calibrated, we only measured the width of the Ca~{\sc ii} K line; see \citet{2004A&A...420..183A} for the full list of stars. We also expanded the sample of metal poor stars with observatons with the ARC echelle spectrograph (ARCES) on the 3.5m telescope at Apache Point Observatory (APO). Observations were made with the standard 1.6\arcsec x 3.2\arcsec slit. Our spectra have a resolution of 31,500, giving us a spectral profile of 7km/s. ARCES covers a wide spectral range of 3,500\AA-10,000\AA.  We also observed some metal-poor giants at Kitt Peak National Observatory on Oct 14-17, 2000. The resolution of the spectrograph used is 18,000. We reduced our spectra with IRAF using a data reduction guide, provided on the APO homepage (www.apo.nmsu.edu/arc35m/Instruments/ARCES/images/echelle\_data\_reduction\_guide.pdf) and the reduction script writen by John Barentine (http://www.apo.nmsu.edu/arc35m/Instruments/ARCES/echellereductionguide.html). We took extreme care to get the wavelength calibration correct, since it is the $\Delta \lambda$ that determines $W$. Our measurements, observation dates, and exposure times are presented in Table \ref{APO}.

\begin{figure}[!ht]
\plotone{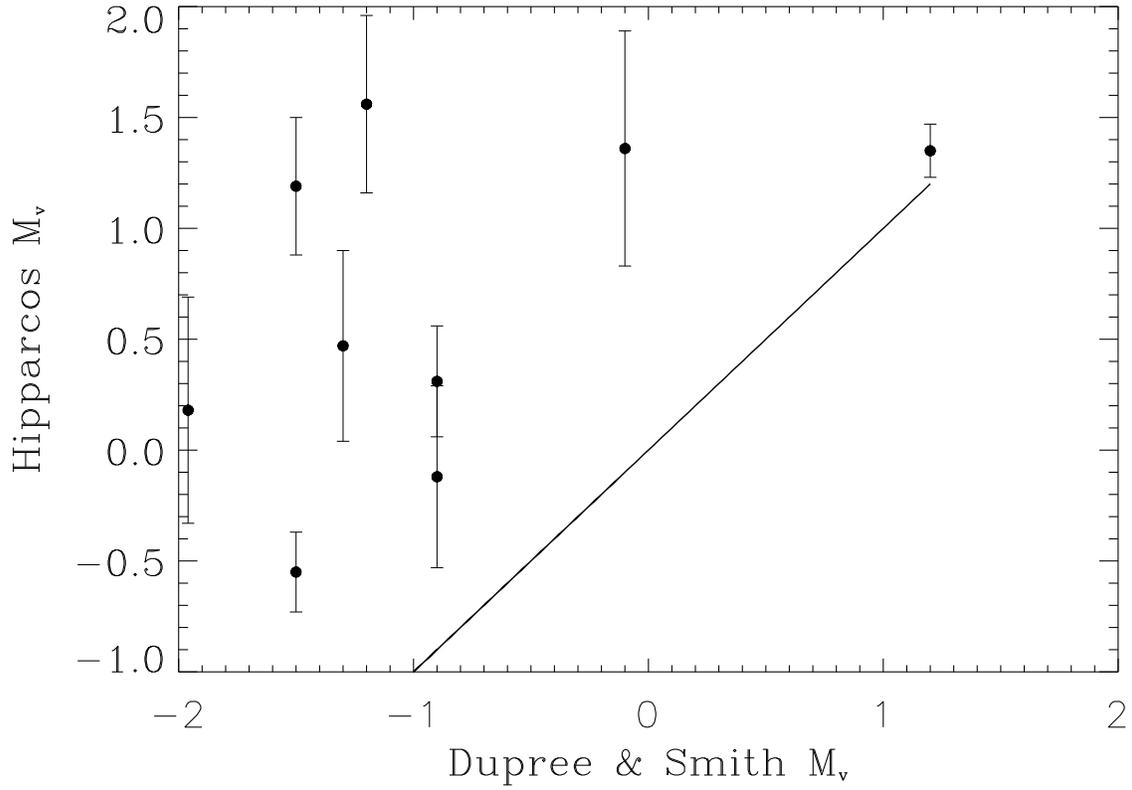}
\caption{The graphical representation of Table \ref{dupree}, demonstrating that the $M_V$ estimated by DS95 do not accuately derive the same $M_V$ from $Hipparcos$ parallaxes. The DS95 absolute magnitudes are systematically brighter than $M_V$ by $Hipparcos$. The solid line shows when both values of $M_V$ agree.  This only includes stars with parallax greater than 3 times the probably error.} 
\label{dupgraph}
\end{figure}

\begin{figure}[!ht]
\plotone{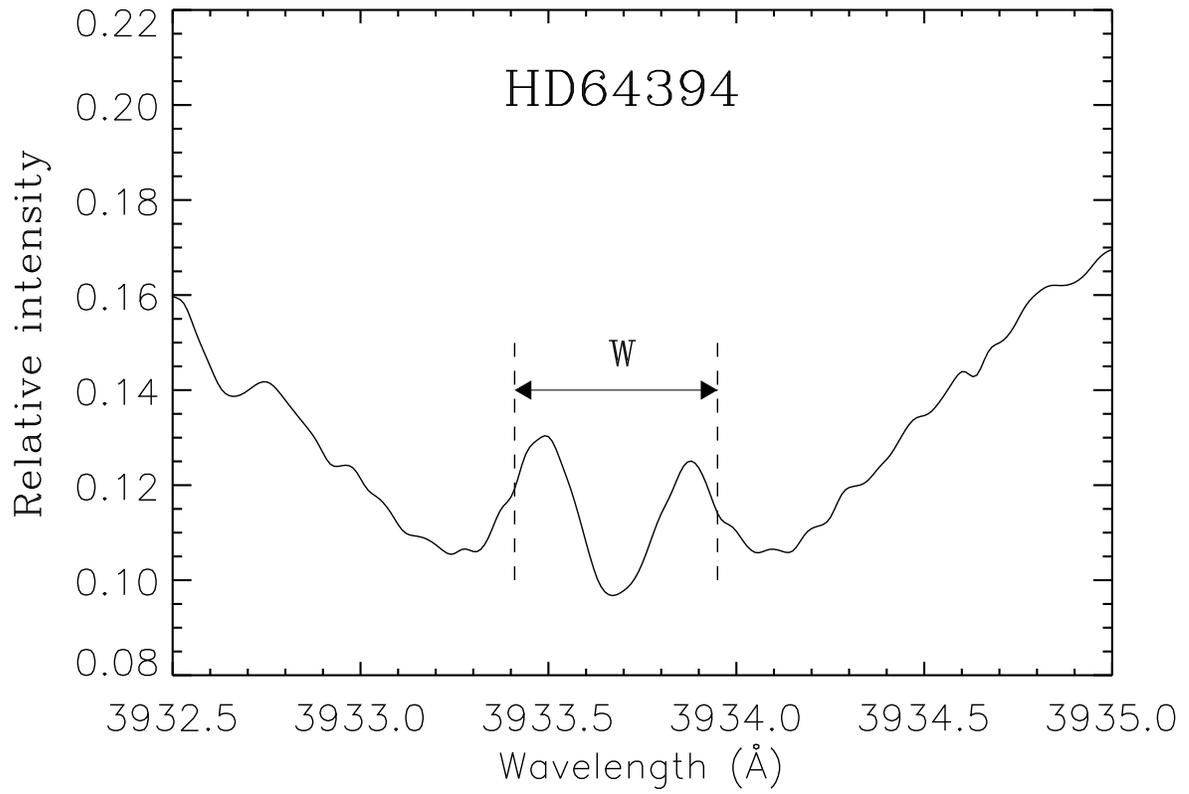}
\caption{The Ca~{\sc ii} K emission profile of HD64394 observed from McDonald. $W$ is measured as the wavelength difference of the half maximum of the blue and red side of the emission profile at the bottom of the K line. W is measured in $km/s$. The instrument profile is then subtracted to get $W_0$, modern instruments have an instrumental profile less than 10$km/s$.}
\label{profile}
\end{figure}

DS95 is of particular interest due to their result of an [Fe/H] dependent WB effect. However, we choose to only include $W_0$ measurements that had a parallax above our 3$\sigma$ criteria. Nine of the twenty-four stars with $W_0$ have accurate parallaxes in $Hipparcos$. However, as we show in Table \ref{dupree} and figure \ref{dupgraph}, these stars' $M_V$ are on average brighter than $M_V$ derived from the 2007 parallaxes. DS95 used M92 to estimate the $M_V$ of all their stars, which have a variety of [Fe/H]. M92 has a luminous and relatively blue RGB due to its low metallicity. When fitting stars of lower metallicity to a bright blue RGB, the stars will have a higher estimated $M_V$ than the star's true $M_V$. We decided to leave out the stars that do not have parallaxes greater than $3\sigma$ due to the inconsistency of using the RGB to estimate $M_V$. Many of these metal poor stars are lost to us since we cannot trust the method used to estimate $M_V$.

\citet{2003yCat..34010997P} and \citet{2006PASP..118..617R} used Gaussian fitting programs to determine $W$ because the stars they observed have strong emission profiles. However, our APO sample is very metal-poor and have small Ca~{\sc ii} K amplitudes is not easily fit with a gaussian profile. Therefore, we chose to measure the profile emission width using the method of \citet{1976ApJ...205..823W}, as shown in Figure \ref{profile}. This method was also used on the full sample of spectra from McDonald \citep{2004A&A...420..183A}. Allende Prieto used the Harlan J. Smith 2.7m telescope and the ESO 1.52m telescope to survey nearby stars with a spectral resolution of $\sim$ 50,000. We compared the WB effect for stars measured by gaussian fitting vs those stars measured using the FWHM and found that there are no systematic differences in $W_0$ space. For the spectra we obtained, we measured the same value of $W$ (within a tolerance) using the two methods. The tolerance becomes negligble after taking the logarithm.

We include studies of globular clusters NGC 2808 and $\omega$ Cen to expand our sample of metal-poor stars, allowing us to compare the WB effect among multiple stellar populations. \citet{2004A&A...413..343C} provided measurements of $\log~W_0$ and $V$ and (m-$M_V$) of RBG stars in NGC 2808. [Fe/H] values of these stars were published by \citet{2004ApJ...610L..25C}. The measurements are shown in Table \ref{ngc2808}. 
Omega Cen has distance modulus of -14.07 (\citealt{1996AJ....112.1487H}, the 2010 edition). The $V$ and [Fe/H] for stars ROA159, 256, 238, 523, WFI321293, and WFI140419 were published by \citet{2011A&A...526A...4V}. The identifier ROA stars are from the catalog by \citet{1966ROAn....2....1W} and the stars identified by WFI were observed with the Wide Field Imaging camera on the 2.2m ESO-MPI telescope in Chile \citep{2000ApJ...534L..83P}. We obtained the spectra used inthese studies and measured $W_0$ of these stars using the same method as APO and McDonald, described above. The $\omega$ Cen stars are presented in Table \ref{omegacen}.

\section{Fitting the Wilson-Bappu effect}

The WB effect was thought to be linear in $\log~W_0$ until \citet{1999PASP..111..335W} demonstrated that the brightest stars have smaller luminosities than their Ca~{\sc ii} widths would suggest. M-dwarfs showed nonlinearity in Mg~{\sc ii} \citet{1990A&A...234..308E} and \citet{2006PASP..118..617R} observed unusual nonlinearity in Ca~{\sc ii} FWHM. We used the stars in \citet{2006PASP..118..617R} to compare against the other stars in our sample. \citet{2006PASP..118..617R} noted that the Ca~{\sc ii} FWHM of those stars were within a few km/s of the intrinsic instumental profile. 

Our full sample shown in Figure \ref{WBE} with a weighted least squares fit of the linear part of the plot (luminosity classes II, III, IV, and bright V's).  Eqn (\ref{fit}) is the line of best fit for stars that fall between $1.4 \leq~\log~W_0~\leq 2.0$ where the errors quoted are the standard deviation of the fit. It is evident that the highest and lowest luminosity class stars do not follow the linear trend of the giants as previously pointed out by \citet{1999PASP..111..335W}. As a group, the supergiants (and perhaps some misclassified luminosity class II stars), follow a different trend than the rest of the giants. There seems to be a step discontinuity at the transition between the M-dwarfs and the K-dwarfs ($\log~W_0~\approx~1.4$) where there is a spread in $M_V$. Accounting for discrepancies in the dwarfs and supergiants is beyond the scope of this study and no higher order polynomial fit was attempted. 

\begin{equation}
M_V = -14.7(\pm0.29)\log~W_0 + 27.0(\pm0.52)
\label{fit}
\end{equation}

\subsection{Low Mass Main Sequence Stars}

Main Sequence stars have a strong and well established relationship between surface gravity, luminosity, and effective temperature. Therefore, we wanted to comapre the expected and observed low luminosity WB effect.  Using the $log~g$, $T_{eff}$, and $M_V$ from the Zero-Age Main-Sequence isochrones by \citet{2000yCat..41410371G} and the relationship ($W_0\sim g^{-2.0}T^{1.1}$) established by \citet{1973A&A....24...79R} and \citet{1974A&A....35...99N} we can compare expected and observed WB effect. The expected and observed low luminosity WB effect is shown in Figure \ref{lowlum}. The WB effect expected from \citet{1973A&A....24...79R} should be linear down to $M_V$ =13. 

Using the conclusions of \citet{1979ApJ...228..509A}, we can speculate that the column depth of M-dwarf chromospheres is a lot smaller, leading to a thinner $W_0$ that we observe. We can use the same argument to say that supergiants have a much thicker chromosphere then models predict, leading to a larger $W_0$ than predicted.

\subsection{Metallicity Dependence}

Because of the large sample and well-behaved linear trend, the luminosity class II,III, IV, and V stars are best suited to examining the effect [Fe/H] has.  Figure \ref{delta} shows residuals from Eqn (\ref{fit}), $\Delta (\log~W_0)$, as a function of [Fe/H]. The averages of $\Delta~\log~W_0$ for the metal-poor stars and the metal-rich stars are $-0.034\pm0.017$ and $-0.048\pm0.042$, respectively. We conclude that on average, the stars with [Fe/H] $\leq$ -1.0 do not deviate significantly from the linear fit. Comparing $\Delta(\log~W_0)$ between the metal-poor stars and metal-rich stars shows that they are drawn from the same distribution. A Kolmogorov-Smirnov test between the metal-poor and metal-rich samples yields a value of 0.27 with a significance of 0.0005, confirming the null hypothesis that the two groups are drawn from the same parent distribution.

\begin{equation}
\Delta \log W_0 = 0.027(\pm 0.004) + 0.0005(\pm 0.004)[Fe/H]
\end{equation}

The lack of [Fe/H] trend disagrees with the conclusions of DS95. The latest $Hipparcos$ $M_V$ differ from the values that they derived using M92 (see Figure \ref{dupgraph}). Using only $M_V$ from $Hipparcos$ parallaxes, the Pearson's r correlation coefficient gives a value of 0.09, indicating that $\Delta\log~W_0$ is uncorrelated with [Fe/H]. The study by \citet{2004A&A...413..343C} compare stars from NGC 2808 with the study by \citet{2003yCat..34010997P} noting that stars in NGC 2808 had a wider spread and tended to be brighter than the fit by \citet{2003yCat..34010997P} would suggest. However, this could just be a sampling effect. Our study shows conclusively that any dependence on [Fe/H] is has a much smaller effect on the WB effect than log g.

We preserve equation \ref{fit} as it stands, adding no term for [Fe/H]. More accurate $M_V$ and a larger sample of $\log~W_0$ will be required for a better WB relation. A re-analysis of the [Fe/H] dependence will be possible after the Gaia satellite measures parallax better than 0.3 mas out to V = 20 and better than 10$\mu as$ at V=15 magnitude. By comparison, Hipparcos measured parallaxes down to 5mas at a limiting magnitude around V=10.

\section{Conclusion}

\begin{figure}[!ht]
\plotone{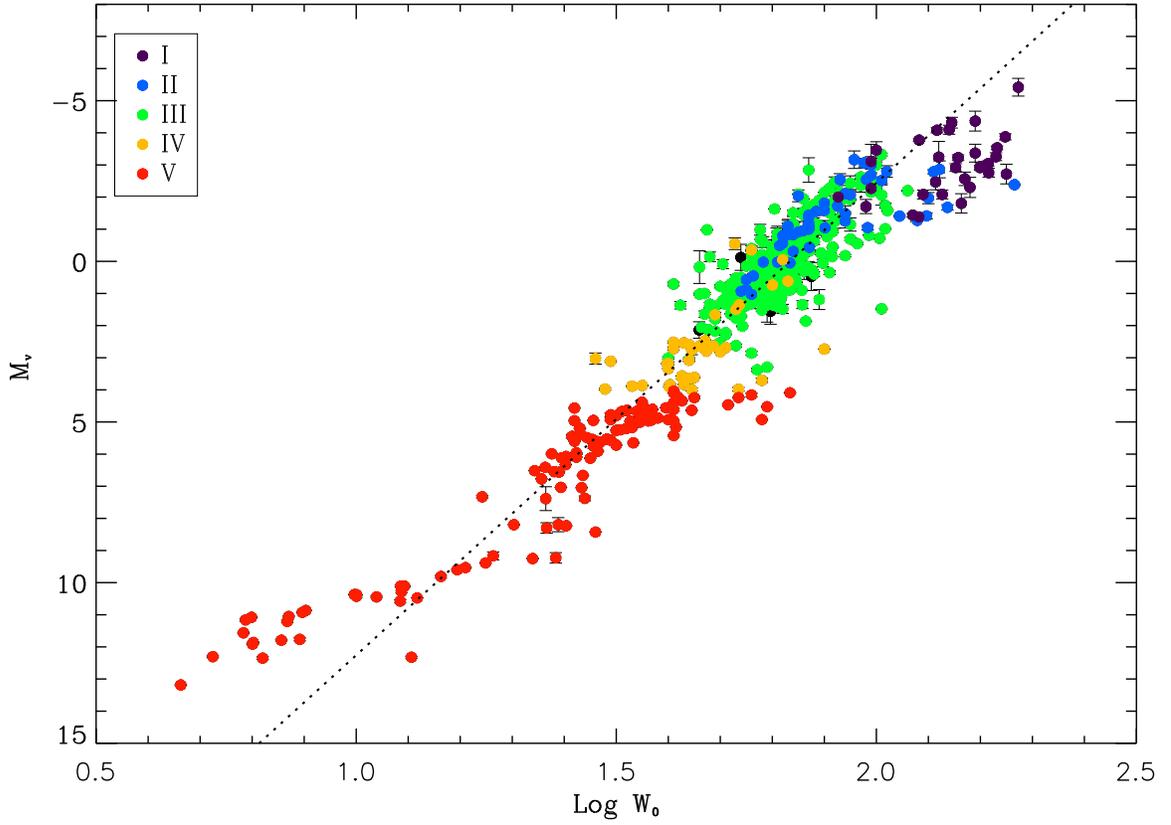}
\caption{The Wilson-Bappu effect shown is broken down by luminosity class. The dotted fit is defined in Equation 1 and is the weighted least squares fit of the linear region between $1.4\leq~\log~W_0~\leq 2.0$. The luminosity class I stars seem to follow a trend that takes them below what Eqn (1) would predict. The M-dwarfs also seem to fall above the fit, having smaller $W_0$ than predicted.}
\label{WBE}
\end{figure}

\begin{figure}[!ht]
\plotone{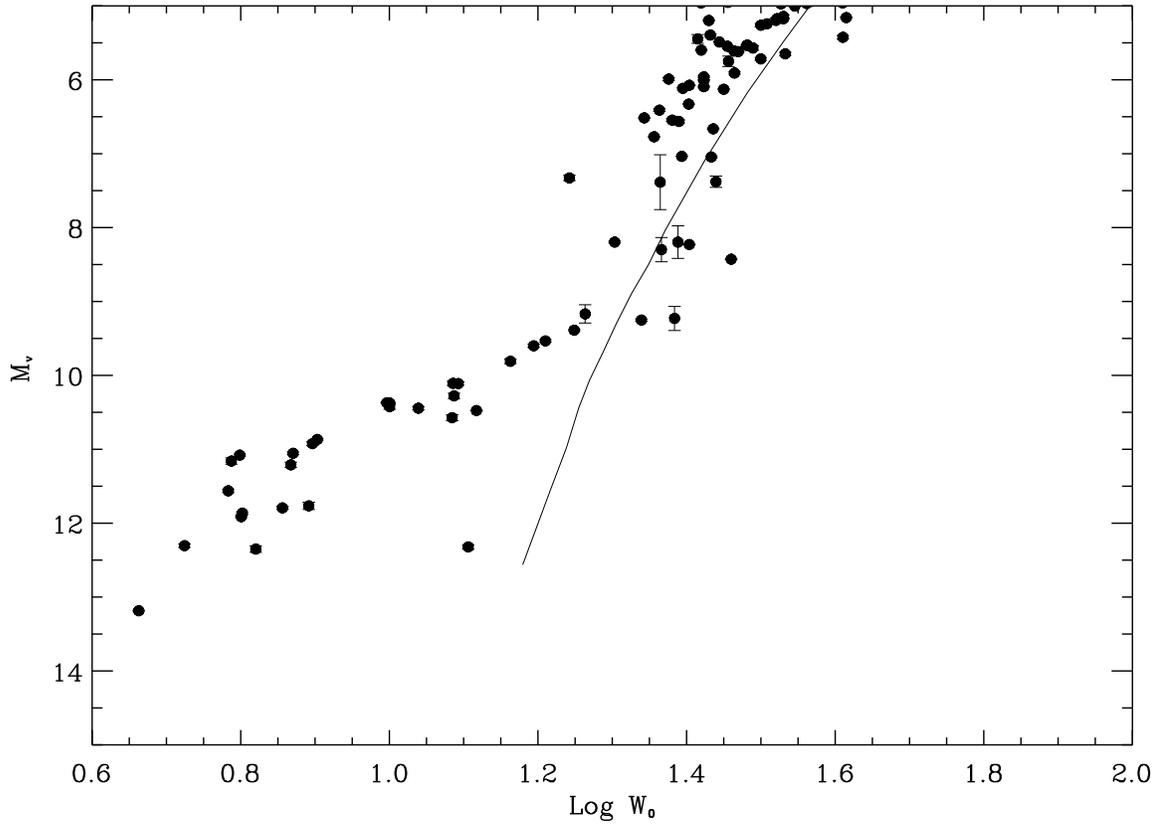}
\caption{The expected Wilson-Bappu effect is shown by the black lines, parameters derived from theoretical isochrones. The data shows that the M-dwarfs have significantly diminished Ca~{\sc ii} widths than predicted. This would indicate a smaller column density in the chromosphere than the stellar models indicate.}
\label{lowlum}
\end{figure}

\begin{figure}[!ht]
\plotone{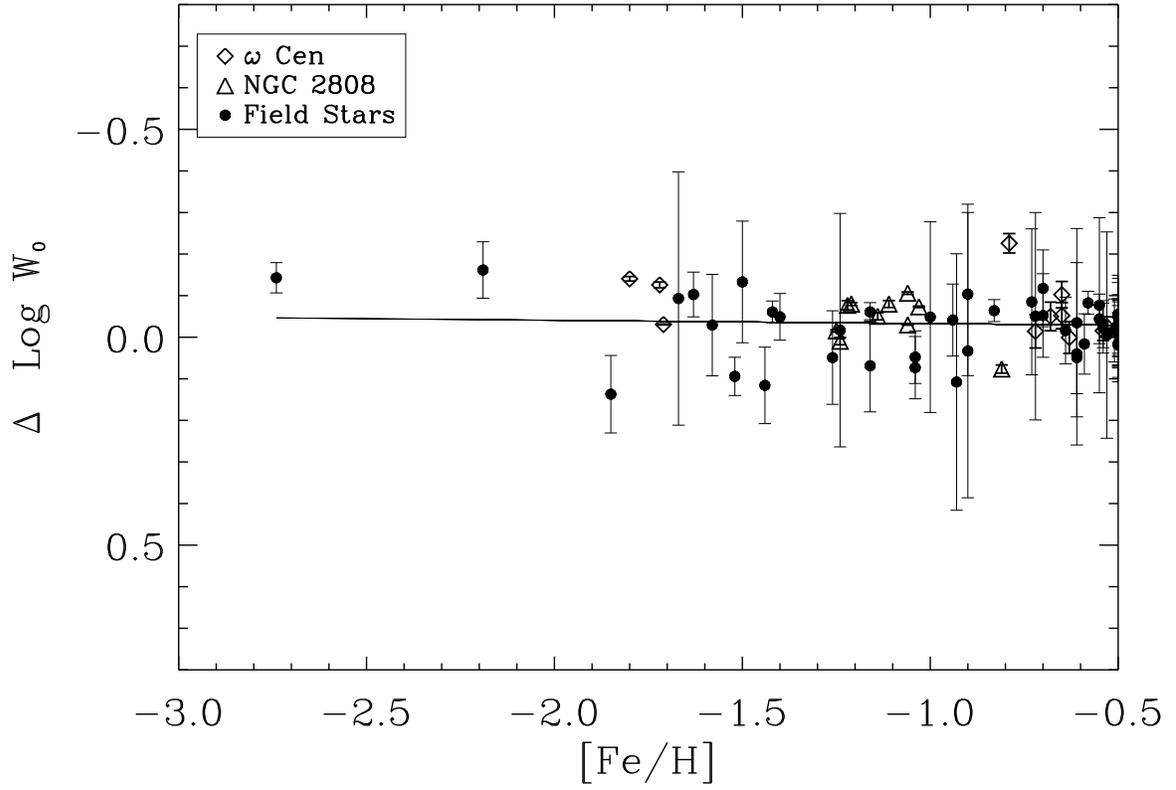}
\caption{The deviations of stars from Eqn (\ref{fit}) ($\log$~$W_0$) are shown as a function of [Fe/H]. The weighted least squares fit is given in equation (2). This plot shows that [Fe/H] has no effect on the deviations from linearity of the giants. This confirms the [Fe/H] dependence predicted by Reimers, Neckel, and Ayres. }
\label{delta}
\end{figure}

We present the largest sample of stars to demonstrate the Wilson-Bappu effect over the largest range of $M_V$ to date.  The main results of this paper are shown in Figs \ref{WBE} and \ref{delta}. Our large sample shows that there is a limited range where the WB effect is linear between $+8 \leq M_V \leq -3$ and $1.35 \leq \log~W_0 \leq 2.0$. Stars deviate from Eq 1 above and below these ranges. To our knowledge, there is no theory of the Ca~{\sc ii} emission width that predicts that the relationship should be linear from $13 \leq M_V\leq -6$ or that a higher-order polynomial would be preferable, the issue of what curve should be drawn through the points is not important. What is important is what we can learn from devitation from the luminosity classes. Ayres (1979) was one of the last developments on width-luminosity relationships like the Wilson-Bappu effect making arguments that increasing column density leads to an increase in the Ca~{\sc ii} width. We can come to the conclusion that the column density of M-dwarf chromospheres are lower than expected while the supergiants have a larger chromospheric column density than expected.

Modern instrumentation such as the $Hipparcos$ satellite and CCD detectors on ground-based telescopes have permitted the WB effect to be extended to about $M_V$=+13. The value of Figure 3. is the recognition of stars that deviate significantly from the mean so that so that their chromospheres may be investigated further in detail. Variations in the intensity of the Ca~{\sc ii} lines in G and K dwarfs have yielded a great deal of information on their rotation and solar-type cycles of varying chromospheric activity \citet{baliunas} .  Our data provides an observing list of stars whose monitoring should produce very interesting results, especially for giants and supergiants of types G, K, and M on their rotations and chromospheric activity variations.

The improved 2007 Hipparcos reduction has allowed us to determine $M_V$ of stars farther away, including more metal-poor stars. High resolution spectra of metal-poor globular cluster giants allowed us to explore the dependence of [Fe/H] to answer some fundamental questions about the WB effect: Is metallicity the cause of the 0.3 magnitude spread, or is it responsible for the deviations of the high and low luminosity trends?. Figure 4 demonstrates clearly the lack of dependence on [Fe/H], as initially asserted by \citet{1957ApJ...125..661W} and theoretically developed by \citet{1979ApJ...228..509A}. 


We thank the Kennilworth fund of the New York Community Trust for support. We also thank the UW McNair program at the University of Washington for their support of T. Gomez.
We are grateful for the efforts of Julie Lutz, Ana Larson, Suzanne Hawley, Erica B\"{o}hm-Vitense, Toby Smith, James R. A. Davenport, and John Barentine who helped with this study and preparation of the manuscript.


\begin{thebibliography}{20}
\expandafter\ifx\csname natexlab\endcsname\relax\def\natexlab#1{#1}\fi

\bibitem[{{Allende Prieto} {et~al.}(2004){Allende Prieto}, {Barklem},
  {Lambert}, \& {Cunha}}]{2004A&A...420..183A}
{Allende Prieto}, C., {Barklem}, P.~S., {Lambert}, D.~L., \& {Cunha}, K. 2004,
  \aap, 420, 183

\bibitem[{{Ambruster}, C.~W. and {Pettersen}, B.~R. and {Sundland}, S.~R.}]{1989A&A...208..198A}
{Ambruster}, C.~W. and {Pettersen}, B.~R. and {Sundland}, S.~R.  1989, \aap, 208, 198


\bibitem[{{Ayres}(1979)}]{1979ApJ...228..509A}
{Ayres}, T.~R. 1979, \apj, 228, 509

\bibitem[{Baliunas} {et~al.}(2005) {Baliunas}, S. L., {Vaughan}, {Hartmann}, {Liller}, {Dupree}]{baliunas}
{Baliunas}, S. L., {Vaughan}, A. H., {Hartmann}, L., {Liller}, W., {Dupree}, A. K.\ 1981 \ ApJ 246, 473

\bibitem[{{Cacciari} {et~al.}(2004){Cacciari}, {Bragaglia}, {Rossetti}, {Fusi
  Pecci}, {Mulas}, {Carretta}, {Gratton}, {Momany}, \&
  {Pasquini}}]{2004A&A...413..343C}
{Cacciari}, C., {Bragaglia}, {Rossetti}, {Fusi
  Pecci}, {Mulas}, {Carretta}, {Gratton}, {Momany} 2004, \aap, 413, 343

\bibitem[{{Carretta} {et~al.}(2004){Carretta}, {Bragaglia}, \&
  {Cacciari}}]{2004ApJ...610L..25C}
{Carretta}, E., {Bragaglia}, A., \& {Cacciari}, C. 2004, \apjl, 610, L25

\bibitem[{{Cayrel de Strobel} {et~al.}(2001){Cayrel de Strobel}, {Soubiran}, \&
  {Ralite}}]{2001yCat.3221....0C}
{Cayrel de Strobel}, G., {Soubiran}, C., \& {Ralite}, N. 2001, VizieR Online
  Data Catalog, 3221, 0

\bibitem[{{Dupree} \& {Smith}(1995)}]{1995AJ....110..405D}
{Dupree}, A.~K., \& {Smith}, G.~H. 1995, \aj, 110, 405


\bibitem[{{Elgaroy} {et~al.} (1990)}] {1990A&A...234..308E} 
{Elgaroy}, O. and {Engvold}, O. and {Carlsson}, M. 1990, \aap, 234,308

\bibitem[Girardi et al.(2000)]{2000yCat..41410371G} Girardi, L., Bressan, 
A., Bertelli, G., 
\& Chiosi, C.\ 2000, VizieR Online Data Catalog, 414, 10371 

\bibitem[{{Glebocki} {et~al.}(1980){Glebocki}, {Musielak}, \&
  {Stawikowski}}]{1980AcA....30..453G}
{Glebocki}, R., {Musielak}, G., \& {Stawikowski}, A. 1980, actaa, 30, 453

\bibitem[{{Harris}(1996)}]{1996AJ....112.1487H}
{Harris}, W.~E. 1996, \aj, 112, 1487

\bibitem[{{Kuhi} (1965)}]{1965PASP...77..253K}
{Kuhi}, L.~V. 1965, \pasp, 77, 253


\bibitem[{{Lastennet} \& {Freire Ferrero}(1994)}]{1994yCat..41080611L}
{Lastennet}, E., \& {Freire Ferrero}, R. 1994, VizieR Online Data Catalog, 410,
  80611

\bibitem[Neckel(1974)]{1974A&A....35...99N} Neckel, H.\ 1974, \aap, 35, 99 

\bibitem[{{Pace} {et~al.}(2003){Pace}, {Pasquini}, \&
  {Ortolani}}]{2003yCat..34010997P}
{Pace}, G., {Pasquini}, L., \& {Ortolani}, S. 2003, VizieR Online Data Catalog,
  340, 10997

\bibitem[{{Pancino} {et~al.}(2000){Pancino}, {Ferraro}, {Bellazzini}, {Piotto},
  \& {Zoccali}}]{2000ApJ...534L..83P}
{Pancino}, E., {Ferraro}, F.~R., {Bellazzini}, M., {Piotto}, G., \& {Zoccali},
  M. 2000, \apjl, 534, L83

\bibitem[{{Perryman} \& {ESA}(1997)}]{1997ESASP1200.....P}
{Perryman}, M.~A.~C., \& {ESA}, eds. 1997, ESA Special Publication, Vol. 1200,
  {The HIPPARCOS and TYCHO catalogues. Astrometric and photometric star
  catalogues derived from the ESA HIPPARCOS Space Astrometry Mission}

\bibitem[{{Rauscher} \& {Marcy}(2006)}]{2006PASP..118..617R}
{Rauscher}, E., \& {Marcy}, G.~W. 2006, \pasp, 118, 617

\bibitem[Reimers(1973)]{1973A&A....24...79R} Reimers, D.\ 1973, \aap, 24, 79 

\bibitem[{{van Leeuwen}(2007)}]{2007A&A...474..653V}
{van Leeuwen}, F. 2007, \aap, 474, 653

\bibitem[{{Vieytes} {et~al.}(2011){Vieytes}, {Mauas}, {Cacciari}, {Origlia}, \&
  {Pancino}}]{2011A&A...526A...4V}
{Vieytes}, M., {Mauas}, P., {Cacciari}, C., {Origlia}, L., \& {Pancino}, E.
  2011, \aap, 526, A4+

\bibitem[{{Wallerstein} {et~al.}(1999){Wallerstein}, {Machado-Pelaez}, \&
  {Gonzalez}}]{1999PASP..111..335W}
{Wallerstein}, G., {Machado-Pelaez}, L., \& {Gonzalez}, G. 1999, \pasp, 111,
  335

\bibitem[{{Wilson}(1976)}]{1976ApJ...205..823W}
{Wilson}, O.~C. 1976, \apj, 205, 823

\bibitem[{{Wilson} \& {Vainu Bappu}(1957)}]{1957ApJ...125..661W}
{Wilson}, O.~C., \& {Vainu Bappu}, M.~K. 1957, \apj, 125, 661

\bibitem[{{Woolley}(1966)}]{1966ROAn....2....1W}
{Woolley}, R.~V.~D.~R. 1966, Royal Observatory Annals, 2, 1

\bibitem[{{Zarro} \& {Rodgers}(1983)}]{1983ApJS...53..815Z}
{Zarro}, D.~M., \& {Rodgers}, A.~W. 1983, \apjs, 53, 815

\end{thebibliography}

\begin{deluxetable}{lccccc}
\tablecolumns{6}
\tablecaption{Metal-Poor Giants from \citeauthor{1995AJ....110..405D}}
\tablehead{
  \colhead{Identifier}& 
  \colhead{$\log$~$W_0$} &
  \colhead{M92 $M_V$} &
  \colhead{$\Pi/\sigma_\Pi$} &
  \colhead{Hipparcos $M_V$} &
  \colhead{[Fe/H]}
 }
\startdata
HD6268&	        1.78&	-1.2&	    & & -2.36\\
HD6833&	        1.83&	-0.9&	8.5 & 0.31$\pm$ 0.25 & -1.04\\
HD8724&	        1.92&	-1.3&	5.5 & 0.47$\pm$0.43 & -1.83\\
HD25532&	1.90&	-0.1&	4.1 & 1.36$\pm$0.53 & -1.26\\
HD26297&	1.83&	-0.9&	    & &  -1.68\\
HD29574&	1.87&	-2.6&	    & &  -1.88\\
HD63791&	1.78&	-0.6&	    & & -1.81\\
HD103036&	2.00&	-2.2&	    & & -1.78\\
HD110184&	1.77&	-2.3&	    & & -2.44\\
HD110281&	1.99&	-2.0&	    & & -1.56\\
HD118055&	1.87&	-1.9&	    & & -1.76\\
HD122563&	1.79&	-1.5&	12.1 & -0.55$\pm$0.18 & -2.74\\
HD122956&	1.80&	-0.9&	5.4  & -0.12$\pm$0.41 & 1.63\\
HIP70199&	1.90&	-2.3&	    &  & -1.99\\
HIP73960&	1.86&	-1.7&	    &  & -1.37\\
HD165195&	1.84&	-1.8&	    &  & -2.24\\
HD166161&	1.85&	-1.2&	5.4 &  1.56$\pm$0.40 & -2.24\\
HD175305&	1.74&	1.2&	17.8 & 1.35$\pm$0.12 & -1.16\\
HD184266&	1.96&	-1.5&	7.0 &  1.19$\pm$0.31 & -1.44\\
HD187111&	1.81&	-1.9&	    & & -1.54\\
HD204543&	1.92&	-0.3&	    & & -1.84\\
HD216143&	1.85&	-1.5&	    & & -2.25\\
HD221170&	1.81&	-1.96&   4.3 & 0.18$\pm$0.51 & -2.19\\
HD232078&	1.90&	-2.0&	    & & -1.54\\
\enddata 
\label{dupree}
\end{deluxetable}

\begin{deluxetable}{lccccc}
\tablecolumns{4}
\tablecaption{APO Observations}
\tablehead{
  \colhead{Identifier}& 
  \colhead{$M_V$}&
  \colhead{$\log$~$W_0$} &
  \colhead{[Fe/H]} & 
  \colhead{Date Obs(UT)}& 
  \colhead{Exposure time}
 }
\startdata
HD6755&        2.14$\pm$0.25& 1.66  & -1.58 & 2010-11-21  1:33:36 & 10min x2\\
&  &  & & 2010-08-21 11:07:27 & 5 min \\
HD6833&        0.31$\pm$0.25& 1.73  & -1.04 & 2010-08-21 10:59:38 & 3 min \\
HD25532&       1.36$\pm$0.52& 1.79  & -1.26 & 2010-04-12  & 5min \\
HD37828\footnote{Observed at Kitt Peak Observaotry}&      -0.98$\pm$0.40& 1.84  & -1.42 & 2000-10-14 & 2hrs \\
HD73394&      -0.34$\pm$0.69& 1.80  & -1.40 & 2011-04-12 03:46:16 & 20 min \\
HD79452&       0.57$\pm$0.15& 1.84  & -0.35 & 2011-04-12 05:41:42 & 3 min \\

HD170737$^1$ &      3.03$\pm$0.17& 1.46  & -0.67 & 2000-10-16 & 1hr 10min \\
HD175305&      1.35$\pm$0.12& 1.85  & -1.44 & 2011-04-12 06:27:34 & 5min x2\\
HD184266$^1$ &      1.19$\pm$0.31& 1.89  & -1.85 & 2000-10-15 &  1hr\\
HD221170&      0.18$\pm$0.50& 1.66  & -2.19 & 2010-08-21 10:09:10 & 5min  
\enddata
\label{APO}
\end{deluxetable}

\begin{deluxetable}{lccc}
\tablecolumns{4}
\tablecaption{$\log W_0$ and [Fe/H] for NGC2808, Identifiers from \citep{2004A&A...413..343C}}
\tablehead{
  \colhead{ID No.}& 
  \colhead{$\log$~$W_0$} &
  \colhead{$M_V$}&
  \colhead{[Fe/H]}
 }
\startdata

37872 &     1.86& -1.940&-1.06\\
46099 &     1.93& -1.849&-1.19\\
46580 &     1.89& -1.900&-1.03\\
47606 &     1.93& -2.154&-1.14\\
48889 &     1.91& -2.249&-1.22\\
50119 &     1.87& -1.704&-1.21\\
50761 &     2.06& -2.200&-0.81\\
51454 &     1.90& -2.144&-1.11\\
51499 &     1.99& -2.155&-1.24\\

\enddata
\label{ngc2808}
\end{deluxetable}

\begin{deluxetable}{lccc}
\tablecolumns{4}
\tablecaption{The $\omega$ Cen Sample}
\tablehead{
  \colhead{Identifier}& 
  \colhead{$\log$~$W_0$} &
  \colhead{V magnitude}&
  \colhead{[Fe/H]}
 }
\startdata
ROA500&	        1.91 &	13.12&	-0.53\\
ROA517&	        1.92  &	13.10&	-0.54\\
WFI139267&	1.73  &	13.09&	-0.79\\
WFI263340&	1.90  &	13.62&	-0.63\\
WFI305654&	1.87  &	13.38&	-0.65\\
ROA159&	        1.89  &	11.98&	-1.72\\
ROA256&	        1.95  &	12.28&	-1.71\\
ROA238&	        1.84  &	12.44&	-1.80\\
ROA523 &	1.82  &	13.39&	-0.65\\
WFI321293&	1.88  &	13.69&	-0.72\\
WFI140419&	1.86  &	13.49&	-0.68\\
\enddata 
\label{omegacen}
\end{deluxetable}

\end{document}